\documentclass[letterpaper, 10 pt, conference]{ieeeconf}  % Comment this line out if you need a4paper

\IEEEoverridecommandlockouts                              % This command is only needed if 
\usepackage{graphics} % for pdf, bitmapped graphics files
\usepackage{epsfig} % for postscript graphics files
\usepackage{mathptmx} % assumes new font selection scheme installed
\usepackage{times} % assumes new font selection scheme installed
\usepackage{amsmath} % assumes amsmath package installed
\usepackage{amssymb}  % assumes amsmath package installed
\usepackage{xcolor}
\usepackage{algorithm}
\usepackage{algpseudocode}
\usepackage{subfigure}
\usepackage{booktabs}
\usepackage{cite}

\title{\LARGE \bf
Modeling Driver Behavior in Speed Advisory Systems: Koopman-based Approach with Online Update}

\author{Mehmet Fatih Ozkan$^{1}$, Jeff Chrstos$^{1}$, Marcello Canova$^{1,2}$ and Stephanie Stockar$^{2}$% <-this % stops a space
\thanks{$^{1}$Center for Automotive Research, The Ohio State University, Columbus, OH, USA.
$^{2}$Department of Mechanical and Aerospace Engineering, The Ohio State University, Columbus, OH, USA. Emails: 
        {\tt\small ozkan.25@osu.edu, chrstos.1@osu.edu, canova.1@osu.edu, stockar.1@osu.edu}}%
}

\begin{document}

\maketitle
\thispagestyle{empty}
\pagestyle{empty}
%%%%%%%%%%%%%%%%%%%%%%%%%%%%%%%%%%%%%%%%%%%%%%%%%%%%%%%%%%%%%%%%%%%%%%%%%%%%%%%%
\begin{abstract}
%In speed advisory systems, where drivers receive optimal speed suggestions based on traffic conditions and energy efficiency goals, the system's effectiveness heavily relies on how drivers respond to the given speed recommendations. Human drivers may not always follow the recommended speed trajectory due to varying driving preferences, external factors, and cognitive biases, which presents a significant challenge to the design of effective and robust driver advisory systems. In this context,  
Accurate driver behavior modeling is essential for improving the interaction and cooperation of the human driver with the driver assistance system. This paper presents a novel approach for modeling the response of human drivers to visual cues provided by a speed advisory system using a Koopman-based method with online updates. The proposed method utilizes the Koopman operator to transform the nonlinear dynamics of driver-speed advisory system interactions into a linear framework, allowing for efficient real-time prediction. An online update mechanism based on Recursive Least Squares (RLS) is integrated into the Koopman-based model to ensure continuous adaptation to changes in driver behavior over time. The model is validated using data collected from a human-in-the-loop driving simulator, capturing diverse driver-specific trajectories. The results demonstrate that the offline learned Koopman-based model can closely predict driver behavior and its accuracy is further enhanced through an online update mechanism with the RLS method.

\end{abstract}

%%%%%%%%%%%%%%%%%%%%%%%%%%%%%%%%%%%%%%%%%%%%%%%%%%%%%%%%%%%%%%%%%%%%%%%%%%%%%%%%
\section{INTRODUCTION}
%In the field of Advanced Driver Assistance Systems (ADAS),  real-time traffic and route information from connectivity and sensing technologies can enable the optimization of vehicle dynamics and powertrain control through look-ahead vehicle energy management systems
In recent years, advanced driver assistance systems (ADAS) such as lane-keeping assist, adaptive cruise control, and automatic emergency braking have been integrated into vehicles to enhance both safety and driver comfort. More recently, there has been a shift towards extending the use of ADAS to improve vehicle energy efficiency. For instance, by incorporating advanced sensing and communication technologies, ADAS can provide real-time speed recommendations to drivers that result in vehicle energy savings \cite{telloni2024evaluating,hong2024energy,rojdestvenskiy2018real}. On the other hand, these recommendations might be disregarded by the driver due to personal preferences, challenges in interpreting audio/visual cues or external factors such as weather and surrounding vehicles, thereby reducing the potential benefits of optimized speed advisories. Furthermore, even when attempting to follow the reference, there is no guarantee that the outcomes will closely match the intended guidance. %This becomes particularly impactful in electric vehicles (EVs), where energy efficiency directly affects the driving range. 
Therefore, it is necessary to develop models that predict human driver responses to ADAS recommendations, improving the system’s effectiveness and efficiency by aligning suggestions with actual driver behavior.\par
Driver behavior modeling has been widely researched in the literature \cite{Nasr}, though understanding driver response to actions or suggestions provided by ADAS is still an emerging field. Existing approaches for driver behavior modeling can be classified into physics-based, theory-based, and data-driven models \cite{ozkan2023interactive}. Physics-based models use mathematical formulations to represent the physical dynamics of driving, such as acceleration and deceleration. Theory-based models aim to capture the underlying cognitive mechanisms and human factors influencing driving behavior, such as attention, risk perception, and decision-making under uncertainty \cite{negash}. Despite the intuitive nature of the physics-based and theory-based approaches, their reliance on fixed model parameters and limited real-time adaptability reduces their effectiveness in capturing the unpredictable behaviors of drivers under real-world conditions, where online learning is essential \cite{Kreutz}. On the other hand, data-driven models offer a more straightforward path to online learning because they are designed to learn from data and adapt quickly and dynamically to new observations \cite{grotto2024online}.\par
Traditional data-driven approaches have relied heavily on machine learning (ML) \cite{abou2020application} and deep learning (DL) \cite{Alkinani} techniques to predict driving patterns over time. These methods typically involve defining a feature matrix to represent driver behavior, often requiring extensive offline training with large datasets and significant computational resources. Furthermore, these strategies can struggle with capturing the nonlinear dynamics of driving behavior \cite{ELAMRANIABOUELASSAD2020103312}. The Koopman operator theory \cite{korda2018linear} presents a compelling alternative to identify complex systems by transforming nonlinear system dynamics into a higher-dimensional linear space with computationally efficient training. Koopman operator theory has been demonstrated to be a promising approach to nonlinear system identification in vehicular applications including eco-driving problems for EVs, autonomous racing, and vehicle platooning \cite{Manzoor}. Although the Koopman method is effective for identifying complex nonlinear systems, it lacks the practicality of real-time updates based on new observations. This can be achieved with online parameter estimation methods like the recursive least squares (RLS) algorithm \cite{isermann2011identification}, as explored in \cite{calderon} and \cite{Sayed}. While these studies demonstrate the use of RLS in updating Koopman-based models, they do not address the specific challenge of modeling dynamic, human-in-the-loop responses to speed advisory systems in real time.
\par
This study proposes a data-driven approach for modeling the human response to a speed advisory system, which addresses the limitations of traditional methods and results in a more adaptable and robust model. First, a Koopman-based driver behavior model is developed to forecast an individual driver’s response to speed advisory cues in a simulator environment. Second, the model is enhanced with the ability to adapt and improve over time by integrating real-time data using the RLS estimator. Unlike previous works, the proposed approach focuses on real-time driver adaptation, making it particularly suited for dynamic driving environments where short-term behavior variations are critical to system performance.
\section{Driver Behavior Learning}
%In this work, the Koopman operator is used to model and identify the behavior of a human driver in response to speed advisory signals. The Koopman operator provides a linear representation of a nonlinear dynamical system by mapping the original state space into a higher-dimensional space of observable functions, allowing the system to be analyzed with linear techniques.

\subsection{Koopman Operator for Controlled Dynamic Systems}
Consider a nonlinear dynamic system described by the state-space equation:
\begin{equation}
    \mathbf{x}_{k+1} = \mathbf{f}(\mathbf{x}_k, \mathbf{u}_k),
\end{equation}
where $\mathbf{x}_k \in \mathbb{R}^n$ is the state of the system at time step $k$, $\mathbf{u}_k \in \mathbb{R}^m$ is the control input, and $\mathbf{f}$ describes the nonlinear dynamics of the system. The Koopman operator, $\mathcal{K}$, governs the evolution of the system's observables, $\mathbf\psi(\mathbf{x})$, rather than the original states, $\mathbf{x}$:
\begin{equation}
    \mathbf{\psi}(\mathbf{x}_{k+1}) = \mathcal{K} \mathbf{\psi}(\mathbf{x}_k),
\end{equation}
where $\mathbf\psi(\mathbf{x}_k)$ is a set of observable basis functions that map the state $\mathbf{x}_k$ into a higher-dimensional space \cite{korda2018linear}. Although the Koopman operator is infinite-dimensional in theory, in practice, it is often approximated by projecting it onto a finite-dimensional subspace $\mathbb{R}^N$, capturing the essential dynamics in a tractable way. The Koopman-based controlled linear predictor then takes the form \cite{korda2018linear}:
\begin{equation}
    \mathbf{\psi}(\mathbf{x}_{k+1}) = A \mathbf{\psi}(\mathbf{x}_k) + B \mathbf{u}_k +\epsilon_k, \ \ \hat{\mathbf{x}}{_k} = C\mathbf{\psi}(\mathbf{x}_k)
        \label{eq:AB}
\end{equation}
where \( A \in \mathbb{R}^{N \times N} \), \( B \in \mathbb{R}^{N \times m} \), $\epsilon$ is an approximation error and \( \hat{\mathbf{x}} \) is the prediction of \( \mathbf{x} \). \( C \in \mathbb{R}^{n \times N} \) projects the lifted state back to the original space (\( \mathbb{R}^N \to \mathbb{R}^n \)).
\subsection{Koopman Operator for Driver Behavior Identification}
\begin{figure}[h!]
    \centering
    \includegraphics[angle=0, scale=0.26]{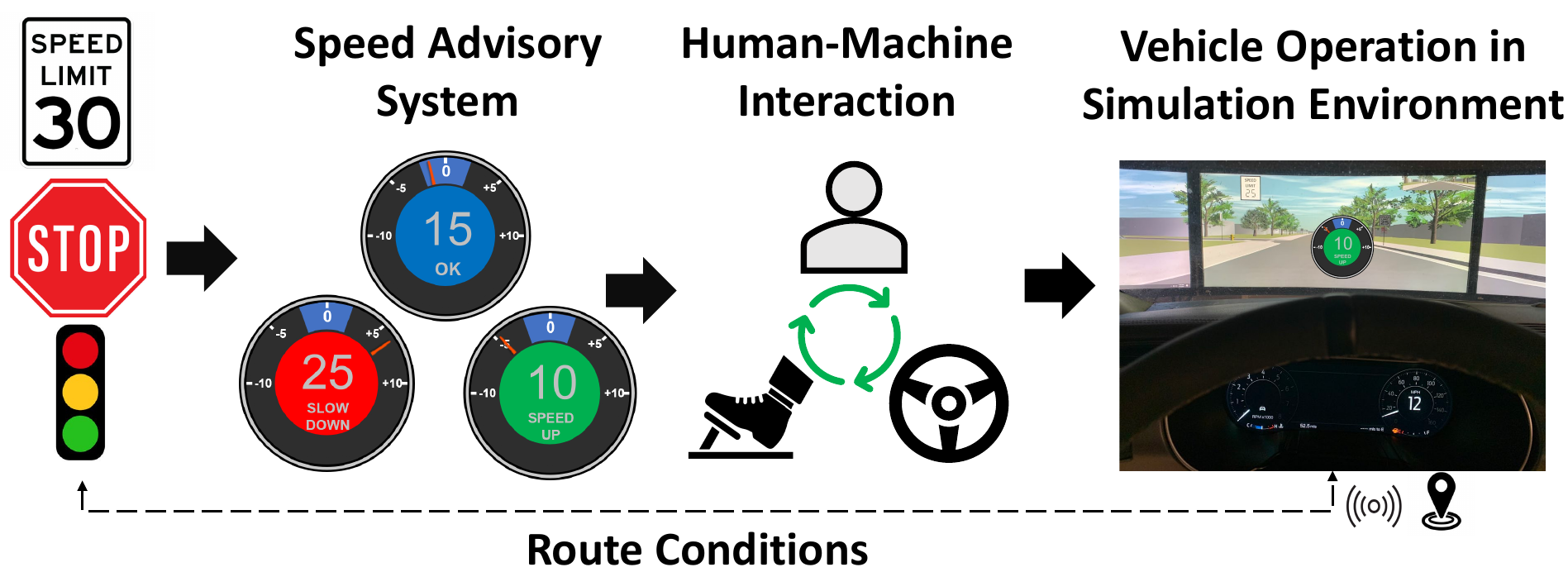}
    \caption{Speed advisory system in the driver-in-the-loop simulator.}
    \label{fig:advisory_sysem}
\end{figure}

In this work, the Koopman operator is applied to identify the nonlinear relationship between driver actions (steering, accelerator, and brake inputs), vehicle longitudinal dynamics, and the speed advisory system in the driver-in-the-loop simulator. The speed advisory system provides real-time target speeds based on GPS and mapping data, displayed visually to the driver via a head-up display, while the vehicle’s updated position and surroundings continuously inform both the driver and the system, as illustrated in Fig. \ref{fig:advisory_sysem}. The system states are the vehicle speed, \( v_k \) and the vehicle traction force, \( F_{tr,k} \). These states are chosen to effectively model the human response to the reference speed \( v_{ref} \), provided by the speed advisory:
\begin{equation}
 \mathbf{x}_k = \begin{bmatrix} v_k, \ Ftr_k \end{bmatrix},\ u_k=v_{ref}
\end{equation}\par
A finite-dimensional approximation of the Koopman operator can be obtained through extended dynamic mode decomposition (EDMD), which uses linear regression on measured data to identify the matrices $A$ and $B$ in Eq. \ref{eq:AB} \cite{williams2015data}.
Given a set of measurements $\{\mathbf{x}_k, \mathbf{x}_{k+1}, u_k\}_{k=1}^{T}$ containing driver-specific trajectories, the state variables are mapped into a set of basis functions, $\mathbf\psi(\mathbf{x}_k)$, forming data matrices $\mathbf{X}$ and $\mathbf{X^+}$ as follows:
\begin{equation}
    \mathbf{X} = \begin{bmatrix} \mathbf\psi(\mathbf{x}_1), \mathbf\psi(\mathbf{x}_2),\cdots, \mathbf\psi(\mathbf{x}_T) \end{bmatrix},
\end{equation}
\begin{equation}
    \mathbf{X^+} = \begin{bmatrix} \mathbf\psi(\mathbf{x}_2),\mathbf\psi(\mathbf{x}_3), \cdots ,\mathbf\psi(\mathbf{x}_{T+1}) \end{bmatrix},
\end{equation}
where $\mathbf{X}$ represents the observables at the current time step, $\mathbf{X^+}$ represents the observables at the next time step. 
Polynomial functions are selected as the problem-specific basis because they effectively capture the nonlinearities in human driver behavior. This is supported by heuristic models, such as the Gipps model \cite{GIPPS1981105} and the enhanced driver model \cite{GUPTA2019574},  which demonstrate that polynomial functions work well for modeling complex, nonlinear human responses in driving environments. In this work, the basis functions are selected up to third-order polynomials of the states:
\begin{equation}
\mathbf{\psi}(\mathbf{x}) = [v_k,Ftr_k,v_kFtr_k,v_k^2,Ftr_k^2,...,v_k^3,Ftr_k^3] 
\label{lifting}
\end{equation}
\par
The input data is collected in the matrix $\mathbf{U}$:
\begin{equation}
    \mathbf{U} = \begin{bmatrix} u_1,u_2 , \cdots, u_T \end{bmatrix}.
\end{equation}\par
The $A$ and $B$ matrices are determined by solving the following linear regression problem:
\begin{equation}
    \min_{A,B} \| \mathbf{X^+} - A \mathbf{X} - B \mathbf{U} \|_F,
\end{equation}
where $\|\cdot\|_F$ is the Frobenius norm. The solution is given by:
\begin{equation}
    \begin{bmatrix} A & B \end{bmatrix} = \mathbf{X^+} \begin{bmatrix} \mathbf{X} \\ \mathbf{U} \end{bmatrix}^\dagger,
\end{equation}
where $\dagger$ denotes the Moore-Penrose pseudoinverse. Since the first two functions in the lifted functions \eqref{lifting} are vehicle velocity and traction force states, the inverse projection to the original state matrix is trivial and given by:
\begin{equation}
\hat{\mathbf{x}}{_k} = C\mathbf{\psi}(\mathbf{x}_k)=
    \left[ \begin{matrix} 1 & 0 & 0 &...&0  \\ 0 & 1 & 0 &...&0 \end{matrix} \right] \mathbf{\psi}(\mathbf{x}_k)
\end{equation}
%By applying EDMD to measurement data, a finite-dimensional approximation of the Koopman operator can be obtained. The matrices $A$ and $B$ represent the linear dynamics of the observables in the lifted space, enabling data-driven identification of the driver’s response to speed advisories. This allows for the prediction of future driver responses based on the identified model.
\subsection{Online Updating of the Koopman Operator}
The response of a human driver to the visual cues from a speed advisory system is inherently dynamic and can change over time due to factors such as fatigue, distraction, or adaptation to road conditions. A static model trained offline may fail to capture these variations, leading to reduced accuracy. To address this limitation, an online updating mechanism for the Koopman-based model that incorporates new real-time measurements is proposed. %By incorporating new measurements in real time, the system can continuously adapt its parameters, ensuring an accurate and up-to-date representation of driver behavior.\par 
One approach to updating the model is to add new measurements to the training dataset \( \mathcal{D} \) and retrain the offline Koopman-based model using the same procedure \cite{calderon}. While this method may work in an offline setting, it becomes impractical in an online context. Continuously adding new data to \( \mathcal{D} \) leads to large memory usage, and as the dataset grows, solving the least squares problem for finding the Koopman matrices \( \mathcal{K} = [A \, B] \) becomes increasingly computationally expensive and difficult to manage. To overcome these limitations, the recursive least square method is used in this work \cite{isermann2011identification}. Instead of storing all historical data, the RLS algorithm updates the matrices \( A \) and \( B \) incrementally based on new observations. This makes it ideal for online applications where real-time performance and memory constraints are critical. \par %RLS method has been recently demonstrated as an efficient approach to adapt the Koopman operator online in \cite{calderon}, \cite{Sayed}. \par
The core idea of the RLS algorithm is to iteratively update the solution to the least squares problem by modifying the Koopman matrices with each set of new measurements. 
The steps for updating the Koopman matrices using the RLS algorithm are outlined in Algorithm \ref{alg:RLS}. The RLS algorithm starts by initializing the Koopman matrices \( A_0 \) and \( B_0 \), along with the covariance matrix \( \mathbf{P}_0 = \lambda^{-1} I \), where \( \lambda \) is the forgetting factor that balances the influence of new and past data. At each time step \( m \), the new lifted measurement \( \mathbf\psi(\mathbf{x}_{m+1}) \) is compared to the predicted value from the current Koopman matrices \( A_m \) and \( B_m \), producing a prediction error \( \mathbf{\epsilon}_m \). The gain matrix \( \mathbf{K}_m \), which determines how much the Koopman matrices should be adjusted, is then updated based on the current lifted measurement  \( \mathbf\psi(\mathbf{x}_m) \) and input \( u_m \), as well as the previous covariance matrix \( \mathbf{P}_{m-1} \). The Koopman matrices \( A_m \) and \( B_m \) are updated using the prediction error and the gain matrix, refining the approximation to account for the new measurements. Finally, after incorporating the latest data, the covariance matrix \( \mathbf{P}_m \) is updated to reflect the reduced uncertainty in the matrices. By iteratively applying these updates, the RLS algorithm provides an efficient mechanism for continuously refining the Koopman matrices without storing all past data, making it well-suited for real-time applications.
\begin{algorithm}[h]
\caption{Online Koopman Model Update with RLS}
\label{alg:RLS}
\begin{algorithmic}[1]
\State \textbf{Input:} Initial Koopman matrices \( A_0 \) and \( B_0 \), forgetting factor \( \lambda \), new state and input measurements $\mathbf{x}_m$ and $\mathbf{u}_m$,  initial covariance matrix \( \mathbf{P}_0 = \lambda^{-1} I \)
\State \textbf{for each time step} \( m = T+1, T+2, \dots, H \)
\State \quad \textit{Compute prediction error}:
\begin{equation}
    \mathbf{\epsilon}_m = \mathbf\psi(\mathbf{x}_{m+1}) - \begin{bmatrix} A_m & B_m \end{bmatrix} \begin{bmatrix} \mathbf\psi(\mathbf{x}_m) \\ u_m \end{bmatrix}
\end{equation}
\State \quad \textbf\textit{Update gain matrix}:
\begin{equation}
    \mathbf{K}_m = \mathbf{P}_{m-1} \begin{bmatrix} \mathbf\psi(\mathbf{x}_m) \\ u_m \end{bmatrix} \left( \lambda + \begin{bmatrix} \mathbf\psi(\mathbf{x}_m) \\ u_m \end{bmatrix}^\top \mathbf{P}_{m-1} \begin{bmatrix} \mathbf\psi(\mathbf{x}_m) \\ u_m \end{bmatrix} \right)^{-1}
\end{equation}
\State \quad \textit{Update Koopman matrices}:
\begin{equation}
    \begin{bmatrix} A_m & B_m \end{bmatrix} = \begin{bmatrix} A_{m-1} & B_{m-1} \end{bmatrix} + \mathbf{K}_m \mathbf{\epsilon}_m^\top
\end{equation}
\State \quad \textit{Update covariance matrix}:
\begin{equation}
    \mathbf{P}_m = \frac{1}{\lambda} \left( \mathbf{P}_{m-1} - \mathbf{K}_m \begin{bmatrix} \mathbf\psi(\mathbf{x}_m) \\ u_m \end{bmatrix}^\top \mathbf{P}_{m-1} \right)
\end{equation}
\State \textbf{end for}
\end{algorithmic}
\end{algorithm}
\section{Case Study}

\subsection{Driver Simulator Setup and Data Collection}
The data for this study were gathered from an experimental campaign using the Vehicle Dynamics Driver-in-the-Loop (VDDiL) simulator at The Ohio State University (OSU) Center for Automotive Research (CAR). The VDDiL simulator combines commercial software with custom-built components to create a versatile platform for driver-in-the-loop experiments. CarSim simulates the vehicle dynamics with high accuracy, while a D-BOX actuation system provides motion in three degrees of freedom (roll, pitch, and heave). The simulator features a SENSO-Wheel SensoDrive that delivers realistic steering feedback. The pedal system includes acceleration and braking controls, with the brake pedal incorporating anti-lock braking system (ABS) feedback, generating pulses when activated. The setup includes a real vehicle cockpit, a three-screen display, and an audio system to enhance immersion. SCANeR Studio by AV Simulation adds environment and traffic simulation capabilities, increasing the system's versatility and level of realism for a range of vehicle development and testing needs. The VDDiL verification against on-road data is presented in \cite{sekar2022assessment}.\par
A vehicle speed advisory system was integrated into the VDDiL simulator. The system displays the target speed for each point of the route on a central screen. The heads-up display includes an inner circle showing the current speed and necessary actions (slow down, maintain speed, or speed up), and an outer circle indicating the difference between the target and current speeds, as illustrated in Fig. \ref{fig:advisory_sysem}. \par
%\begin{figure}[h!]
%    \centering
%    \includegraphics[angle=0, scale=0.60]{driver_view.pdf}
%    \caption{A scene from the driver's interaction with the speed advisory in VDDiL simulator.}
%    \label{fig:advisory}
%\end{figure}
Driver-specific data were collected on a 7.4 km simulated urban route, designed to replicate real-world commuting conditions in Columbus, OH. A human subject study was conducted on 18 subjects with diverse driving experience, and driving style. After a brief training session to familiarize participants with the simulation, each participant was instructed to drive the route while following the speed advisory cues as closely as possible.
%\begin{figure}[h!]
 %   \centering
  %  \includegraphics[angle=0, scale=0.38]{schematic_ACC 25_route_V2.pdf}
   % \caption{A view of the urban route map and simulated environment.}
    %\label{fig:route}
%\end{figure}
%
%\begin{figure}[h!]
%    \centering
%    \includegraphics[angle=0, scale=0.27]{schematic_ACC 25_DIL_V2.pdf}
%    \caption{Driver data collection process in the VDDiL simulator.}
%    \label{fig:schematic}
%\end{figure}
\subsection{Advisory Speed Generation}
The speed advisory profile provided to human drivers was generated offline using a dynamic programming (DP)-based eco-driving strategy. This strategy is formulated as a nonlinear spatiotemporal trajectory optimization problem, aiming to balance travel time and fuel consumption by jointly optimizing the vehicle and powertrain dynamics of a 2017 Chrysler Pacifica plug-in hybrid EV (PHEV) along a specified route \cite{ozkan2024optimizing}.\par
\begin{figure}[ht]
    \centering
    \includegraphics[angle=0, scale=0.56]{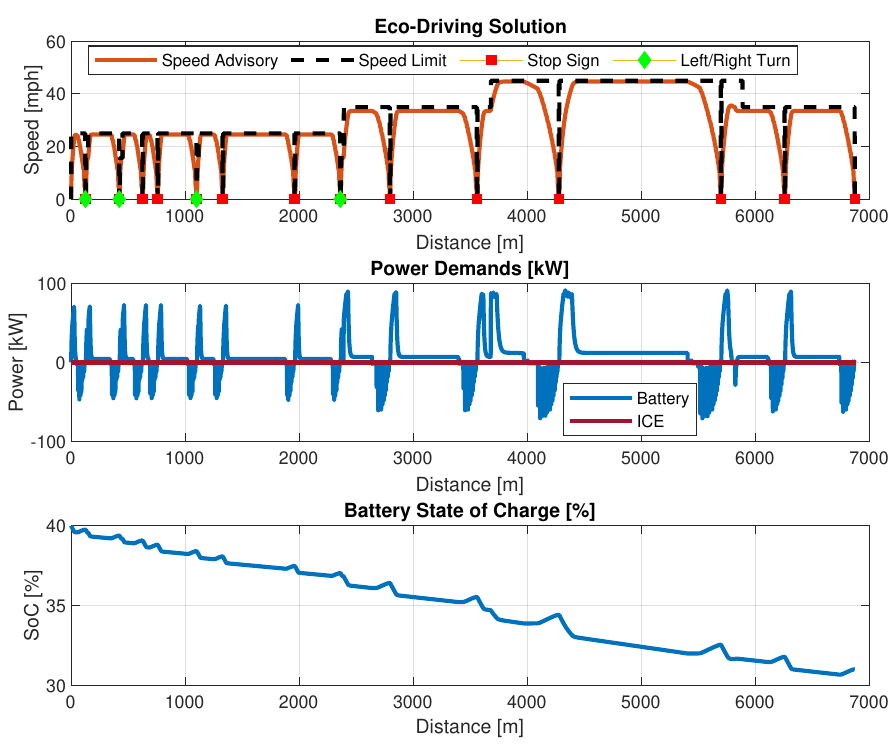}
    \caption{Speed profiles, power demands, and battery state of charge profiles provided by the optimal solution in the speed advisory system.}
    \label{fig:eco}
\end{figure}

%The eco-driving strategy is formulated as a nonlinear spatio-temporal trajectory optimization problem, where the state dynamics are defined as:
%\begin{equation}
%    x_{s+1} = f(x_s, u_s), \quad s = 1, \ldots, N
%\end{equation}
%where $x_s \in X_s \subset \mathbb{R}^n$ represents the states and $u_s %\in U_s \subset \mathbb{R}^m$ represents the control inputs, with $N$ %denoting the total number of itinerary steps along the defined route. 
In this formulation, the state vector $x_s$ at distance step $s$ includes vehicle velocity $v_{\text{veh},s}$ and battery state of charge (SoC) $\xi_s$. The control input vector $u_s$ consists of the internal combustion engine (ICE) on/off decision flag, ${ICE}_{\text{ON/OFF},s}$, which determines the vehicle's operation mode (HEV or EV) and the vehicle acceleration $a_s$. % The discrete state dynamics are expressed as:
% \begin{equation}
%     v_{\text{veh},s+1} = \sqrt{v_{\text{veh},s}^2 + 2\Delta da_s}, \     \xi_{s+1} = \xi_s - \frac{\Delta d}{v_{\text{veh},s} } \frac{I_{\text{batt},s}}{ C_{\text{nom}}}
%     \label{eco_1}
% \end{equation}
% where $\Delta d$ is the discretization step size, set to 2 m, $I_{\text{batt},s}$ and $v_{\text{veh},s}$ represent the average battery current and average vehicle velocity over a distance step, respectively, and $C_{\text{nom}}$ is the nominal battery capacity.
Given that route information, including speed limits, stop signs, and traffic light positions is available in advance from an advanced navigation system, the full-route eco-driving problem over $N$ distance steps is formulated as:
\begin{equation}
    J_s^\ast = \min_{\{\mu_s\}_{s=1}^N} \left\{ c_{N+1}(x_{N+1}) + \sum_{s=1}^{N} c_s(x_s, \mu_s(x_s, u_s)) \right\}
\end{equation}
\begin{equation}
    c_s(x_s, \mu_s(x_s, u_s)) = \left(\gamma \frac{\dot{m}_{\text{eqf},s}}{\dot{m}_{\text{norm}}} + (1 - \gamma) \right) \Delta t_s
\end{equation}
subject to the constraints:
\begin{equation}
    v_{\text{veh}}^{\min} \leq v_{\text{veh},s} \leq v_{\text{veh}}^{\max},\     \xi^{\min} \leq \xi_s \leq \xi^{\max}
\end{equation}
\begin{equation}
    \text{ICE}_{\text{ON/OFF},s} \in \{0, 1\},\    a^{\min} \leq a_s \leq a^{\max}
\end{equation}
\begin{equation}
    v_{\text{veh},1} = v_{\text{veh},1}^{\min},\ \xi_1 = \text{SoC}_0,\    \xi_{N+1} > 26\%
    \label{eco_last}
\end{equation}
where $u_s: X \times U$ represents the admissible control policy at distance step $s$, $c_s: X \times U \rightarrow \mathbb{R}$ is the stage cost, defined as the weighted sum of equivalent fuel consumption and travel time, and $c_{N+1}$ is the terminal cost. The parameter $\gamma$ is the trade-off factor between fuel efficiency and travel time, ranging from 0 to 1. Additionally, $\dot{m}_{\text{eqf},s}$ denotes the equivalent fuel consumption rate, while $\dot{m}_{\text{norm}}$ is a normalization factor. $\Delta t_s$ refers to the travel time over a given distance step. The variables $v_{\text{veh}}^{\text{min}}$ and $v_{\text{veh}}^{\text{max}}$ represent the minimum and maximum speed limits, respectively, and $a^{\text{min}}$ and $a^{\text{max}}$ are the bounds for longitudinal acceleration. In this study, the trade-off factor $\gamma$ was set to 0.5, and the initial state of charge of the battery $\xi_1$ was 40\%. Readers are referred to \cite{ozkan2024optimizing} for further details of the eco-driving strategy.
%\subsection{Advisory Speed Assessment}
 Fig. \ref{fig:eco} illustrates the solution to the eco-driving optimization problem described for an initial SoC of 40\%. This solution was used as the reference velocity for the speed advisory system in the human subject study. The energy-optimal solution results in smooth speed trajectories, minimizing unnecessary braking and acceleration events. This behavior is notably consistent when approaching stop signs and turns. The optimal solution also recommends maintaining a steady cruising speed and optimizing energy efficiency while adhering to speed limits. Additionally, the eco-driving solution maximizes the use of EV mode, avoids activating the ICE, and utilizes regenerative braking early when slowing down, further enhancing energy efficiency.

\section{Results and Discussion}
%\subsection{Driver Behavior Model Evaluation}
\subsection {Offline Koopman-based Driver Behavior Model Development}
\begin{figure}[ht]
    \centering
    \includegraphics[angle=0, scale=0.55]{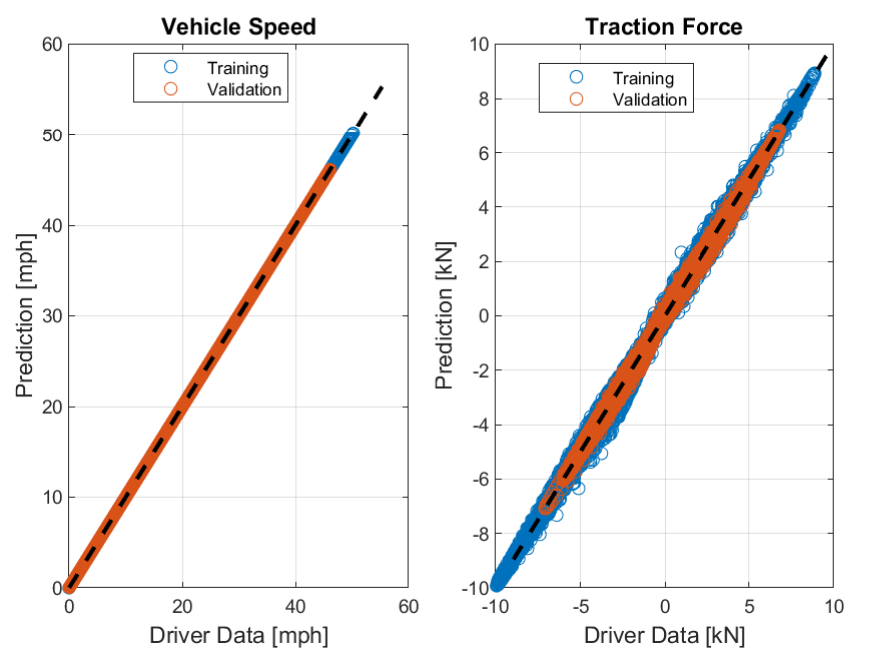}
    \caption{Training and validation results of the Koopman-based driver model.}
    \label{fig:training}
\end{figure}

The Koopman-based driver model is developed using 18 driver-specific trajectories, totaling approximately 3.4 hours of data collected at a 40 Hz sample rate from the VDDiL simulator. These trajectories were cascaded and then split into training (80\%), validation (10\%), and testing (10\%) datasets for the Koopman-based model. Fig. \ref{fig:training} shows the model's performance during the training and validation phases. The results indicate that the Koopman-based model can closely match the experimental data for vehicle speed and traction force states. This close alignment between predicted and actual data suggests that the model effectively captures the driver-vehicle interaction dynamics during training and validation.\par
\begin{figure}[ht]
    \centering
    \includegraphics[angle=0, scale=0.58]{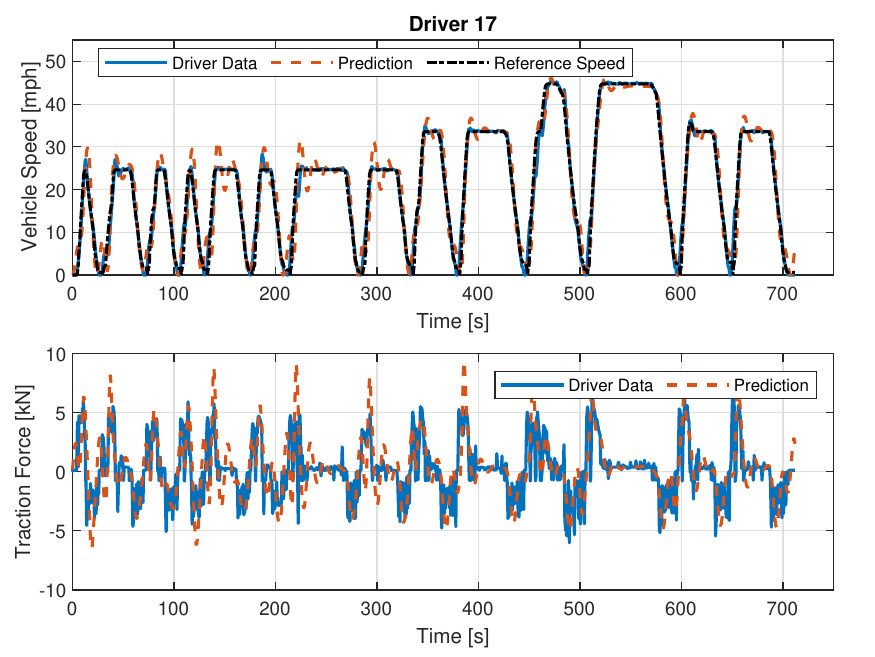}
    \caption{Testing results of the Koopman-based driver model for Driver 17 (without online update).}
    \label{fig:Driver17}
\end{figure}

\begin{figure}[ht]
    \centering
    \includegraphics[angle=0, scale=0.58]{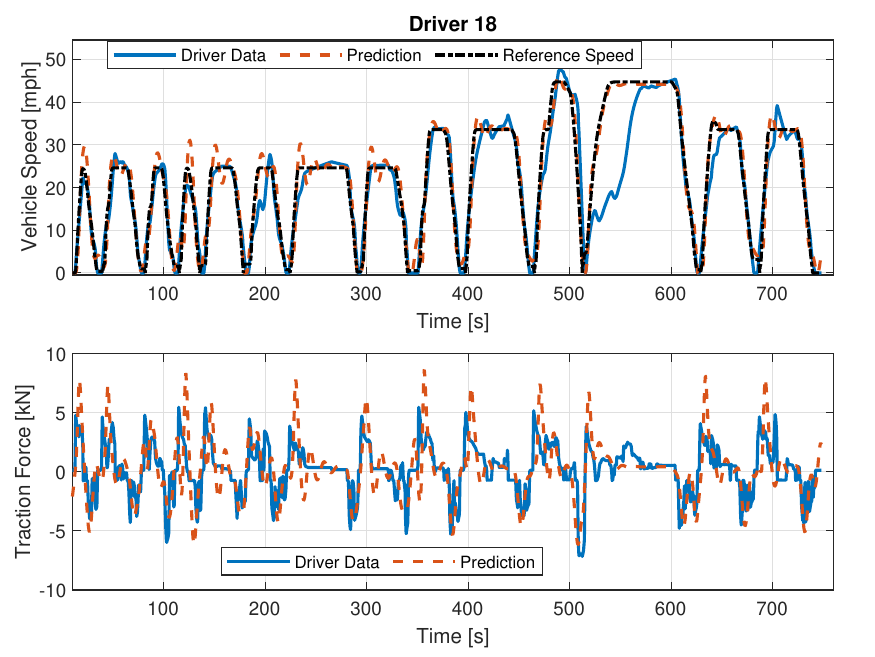}
    \caption{Testing results of the Koopman-based driver model for Driver 18 (without online update).}
    \label{fig:Driver18}
\end{figure}

For the testing phase, two distinct driver-specific trajectories from the data (Driver 17 and Driver 18) are examined. The model is exercised by initializing the vehicle speed and traction force at the beginning of the route and using the reference speed as the model input. Fig. \ref{fig:Driver17} and Fig. \ref{fig:Driver18} present the testing results of Drivers 17 and 18. The trained Koopman-based model demonstrates a strong ability to capture the drivers' speed and traction force trajectories, reflecting its effectiveness in replicating observed driving behaviors. This is noteworthy given that the model states were initialized only at the start of the testing process. However, there are notable instances, particularly with Driver 18, where the model's predictions do not align as closely with the actual data. One such instance occurs between the 515 and 630 seconds in Fig. \ref{fig:Driver18}, where the driver exhibits uncertain behavior, which is possibly due to inattention to the speed advisory input. %This discrepancy highlights the challenges of accurately modeling complex driving dynamics, especially when drivers may not consistently adhere to advisory cues. 
\par
\begin{figure}[h!]
\centering
\subfigure{
\includegraphics[angle=0, scale=0.45]{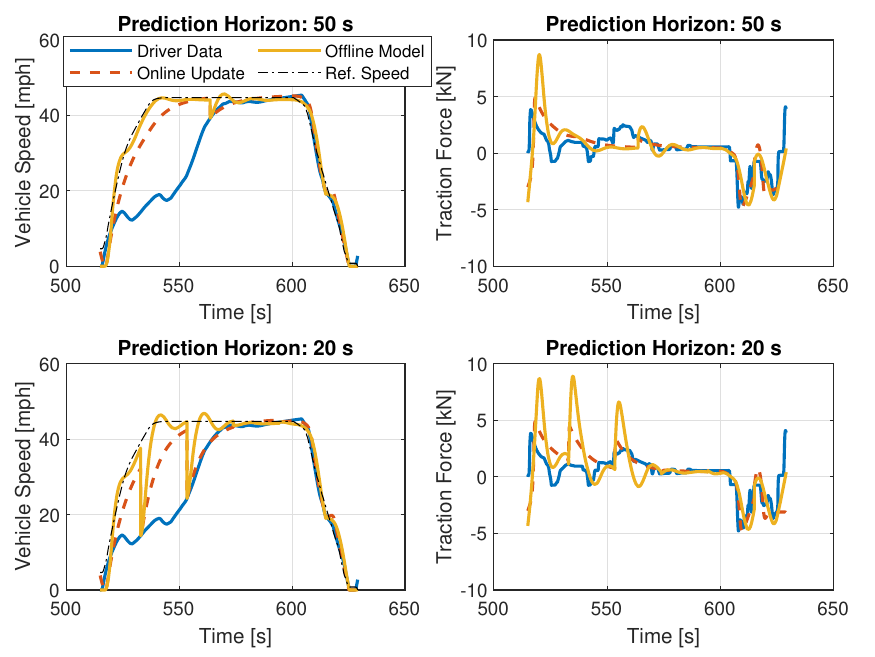}  
}
\subfigure{
\label{fig:1}
\includegraphics[angle=0, scale=0.45]{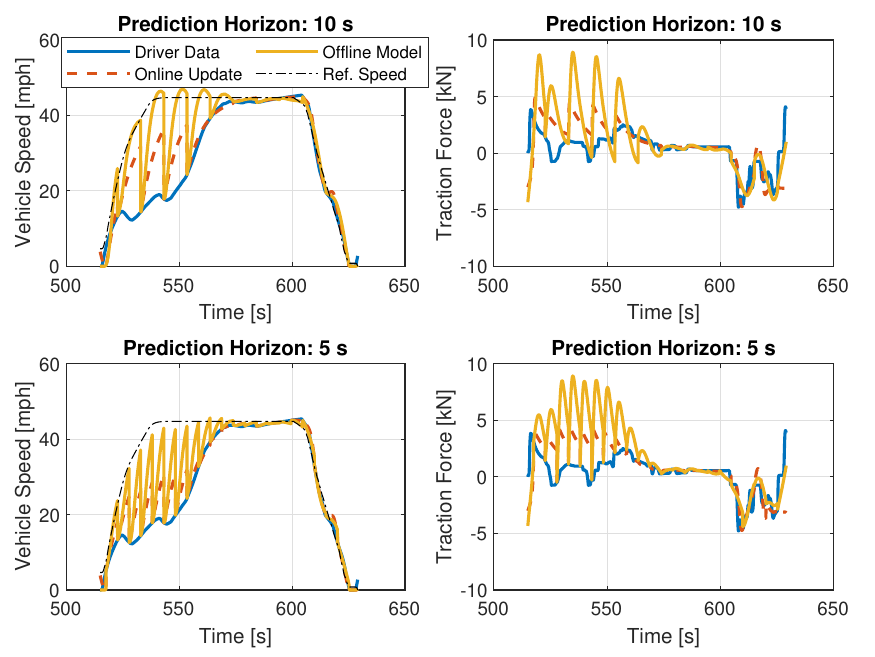}    %fig1.eps has one
}
\caption{Comparison between the offline learned and online updated Koopman-based model between 515 and
630 seconds of Driver 18 under different prediction time horizons.}
\label{fig:online_results}
\end{figure}
\subsection {Online Driver Behavior Learning Analysis}
The proposed RLS-based online update mechanism for the Koopman-based model is evaluated to address discrepancies in the driver's behavior when following the advisory cues. The RLS method is evaluated for Driver 18, between 515 and 630 seconds, focusing on the segment of the route with the largest modeling error. The online learning process is performed with the forgetting factor $\lambda=0.9$ and the Koopman operator is updated every 1 s of the driver observations during that period. The online Koopman-based model is compared against the offline learned Koopman-based model under different prediction time horizon lengths. In this analysis, the initial conditions for the velocity and traction force states are set using the measured values at the end of each prediction horizon.
The result of this comparison is shown in 
Fig.  \ref{fig:online_results}. %illustrates predicted velocity and traction force trajectories of the offline learned and online Koopman-based model between 515 and 630 seconds of Driver 18 under different prediction time horizons. 
The online Koopman-based model significantly outperforms the offline version in predicting human driver behavior. This improvement is due to its ability to make real-time changes, which cannot be achieved by merely reinitializing the offline model. By integrating the RLS approach, the online model continuously updates to capture short-term variations in behavior, leading to substantially improved prediction accuracy compared to the offline model. \par

To compare the prediction accuracy performance of the online and offline Koopman approaches, the Root Mean Square Error (RMSE) values between predicted and actual trajectories are calculated and summarized in Tables \ref{speed_rmse} and  \ref{traction_rmse}. The online Koopman-based model reduces the speed prediction error by 17\% to 49\% and the traction force prediction error by 7\% to 42\% compared to the offline model. The most significant improvements are consistently observed over the shortest prediction horizons for both metrics. This is primarily due to the online model's ability to respond to changes in driver behavior in real time, enabling it to capture immediate shifts and short-term dynamics that the offline model may miss. This adaptability enhances prediction accuracy, especially over shorter horizons, where rapid changes in Driver 18's behavior are more frequent.\par

The total computation time for the online learning process of the Koopman-based model is compared to that of the retrained offline model with new measurements. As shown in Table \ref{computation}, online learning with RLS significantly reduces computation time compared to retraining the Koopman model. This reduction is due to the continuous addition of data in the offline model, which increases the dataset size and makes solving the least squares problem for the Koopman matrices computationally intensive. In contrast, the RLS algorithm iteratively updates the Koopman matrices with new measurements without storing the entire dataset, making it far more practical for online learning. %This improvement is primarily due to the offline model's inability to adapt to real-time changes in driver behavior during human-speed advisory interactions. %In contrast, the online approach enhances prediction performance by updating the model parameters with the observed behavior of Driver 18 using the RLS method. The results collectively demonstrate that the precision and online adaptability of the Koopman method make it a promising approach for modeling driver response given the speed advisory systems.

\begin{table}[h]
    \centering
    \caption{Velocity RMSE  [mph] results of the Koopman-based Model with and without online update.}
    \begin{tabular}{ccccc}
        \toprule
        Prediction Horizon [s]  & 50s & 20s & 10s & 5s \\
        \midrule
        Offline  Model & 12.57   & 11.58  & 10.78   & 7.22\\
        \hline
        Online  Model&  10.45   & 7.98   & 5.93   & 3.69 \\
            \bottomrule
    \end{tabular}
    \label{speed_rmse}
\end{table}
\begin{table}[h]
    \centering
    \caption{Traction Force RMSE [kN] results of the Koopman-based Model with and without online update.}
    \begin{tabular}{ccccc}
        \toprule
        Prediction Horizon [s]  &50s& 20s&10s&5s \\
        \midrule
        Offline Model & 1.58&2.10&2.61&3.16\\
        \hline
        Online Model&1.46&1.54&1.67&1.83 \\
            \bottomrule
    \end{tabular}
    \label{traction_rmse}
\end{table}

\begin{table}[h]
    \centering
    \caption{Computation time [s] comparison of Offline Koopman-based Model with Retraining and Online Update with RLS.}
    \begin{tabular}{ccccc}
        \toprule
        Prediction Horizon [s]  &50s& 20s&10s&5s \\
        \midrule
        Offline Model with Retraining & 4.48&4.71&4.37&4.33\\
        \hline
        Online Model&0.04&0.04&0.03&0.04 \\
            \bottomrule
    \end{tabular}
    \label{computation}
\end{table}

\section{Conclusions and Future Work}
In this paper, a novel method for modeling driver behavior in response to speed advisory systems with real-time adaptability is established. A Koopman-based driver behavior model is developed using data from a comprehensive human subject study with a driver-in-the-loop simulator. Results show that the developed model can effectively capture the dynamics of the driver-speed advisory interaction across different drivers. The inclusion of an online updating mechanism enabled by the RLS algorithm further improves prediction accuracy, especially in cases where the driver's behavior deviates from the recommended speed trajectory. Future work will explore personalized speed-advisory strategies integrated with the proposed driver behavior model to improve human-speed advisory interactions.
\addtolength{\textheight}{-12cm}   
\section*{ACKNOWLEDGMENT}
The authors are grateful to the OSU Center for Automotive Research for providing funding in support of this work and the authors also would like to acknowledge Stellantis for the inspiring discussions and helpful feedback that contributed to the research presented in this paper.
\section*{REFERENCES} 
\begingroup
\renewcommand{\section}[2]{} % This suppresses the automatic section title
\bibliographystyle{ieeetr}
\bibliography{mybibfile}
\endgroup

\end{document}